\documentclass[aps,prl,twocolumn,superscriptaddress,fleqn,a4paper]{revtex4-1}
\usepackage[ngerman,english]{babel}
\usepackage[utf8]{inputenc}   
\usepackage{xspace}
\usepackage{amsmath,tensor}
\usepackage{array,color}
\usepackage{relsize,exscale}  
\usepackage{textcase}	
\usepackage{soul,color}		
\usepackage{calc} 
\usepackage{makeidx,showidx}
\usepackage{graphicx}
\usepackage{wrapfig}
\usepackage{float}
\usepackage{textcomp}
\usepackage{siunitx}
\usepackage[version=3]{mhchem}
\usepackage{enumitem}
\usepackage{fancyref}
\usepackage[hidelinks=true]{hyperref}
\sethlcolor{green}
\setulcolor{red}
\usepackage{lineno}			

\newcommand{\erww} [1] {\ensuremath{\langle {#1} \rangle}}

\newcommand{\lsco} {{La$_{2-x}$Sr$_x$CuO$_4$}\@\xspace}

\newcommand{\hgryb} {{HgBa$_{2}$CuO$_{4+\delta}$}\@\xspace}

\newcommand{\ybcoF} {$\ce{YBa2Cu3O_{7}}$\@\xspace}
\newcommand{\ybco} {$\ce{YBa2Cu3O_{6+y}}$\@\xspace}

\newcommand{\ybcoE} {$\ce{YBa2Cu4O8}$\@\xspace}

\newcommand{\tind} {{$T$\xspace independent}\@\xspace}
\newcommand{\tc} {\ensuremath{T_{\rm c}}\@\xspace}
\newcommand{\cperp}{\ensuremath{c \bot B_0}\@\xspace}
\newcommand{\cpara}{\ensuremath{{c\parallel\xspace B_0}}\@\xspace}

\newcommand{\temp} {{\ensuremath{T}}\@\xspace}

\newcounter{exex}[section]

\makeatletter\newcommand\listofexamples{\section*{List of Examples}\@starttoc{xmp}}
	\newcommand\l@example[2]{\par\noindent#1~\textit{#2}\par}
\makeatother






\makeatletter
\renewcommand\subsection{\@startsection 
{subsection}{3}{0mm}
{-\baselineskip}
{0.5\baselineskip}
{\centering \textbf }}
\makeatother

\makeatletter
\renewcommand\subsubsection{\@startsection 
{subsubsection}{3}{0mm}
{-\baselineskip}
{0.5\baselineskip}
{\centering  }}
\makeatother

\makeindex

\begin{document}
\title{Properties of the electronic fluid of superconducting cuprates from $^{63}$Cu NMR shift and relaxation}

\date{\today}
\author{Marija Avramovska}
\author{Danica Pavi\'cevi\'c}
\author{Jürgen Haase}
\affiliation{University of Leipzig, Felix Bloch Institute for Solid State Physics, Linn\'estr. 5, 04103 Leipzig, Germany}

\begin{abstract}
Nuclear magnetic resonance (NMR) provides local, bulk information about the electronic properties of materials, and it has been influential for theory of high-temperature superconducting cuprates. Importantly, NMR found early that nuclear relaxation is much faster than what one expects from coupling to fermionic excitations above the critical temperature for superconductivity ($T_{\rm c}$), i.e. what one estimates from the Knight shift with the Korringa law. As a consequence, special electronic spin fluctuations have been invoked. Here, based on literature relaxation data it is shown that the electronic excitations, to which the nuclei couple with a material and doping dependent anisotropy, are rather ubiquitous and Fermi liquid-like. A suppressed NMR spin shift rather than an enhanced relaxation leads to the failure of the Korringa law for most materials. Shift and relaxation below $T_{\rm c}$ support the view of suppressed shifts, as well. A simple model of two coupled electronic spin components, one with $3d(x^2-y^2)$ orbital symmetry and the other with an isotropic $s$-like interaction can explain the data. The coupling between the two components is found to be negative, and it must be related to the pseudogap behavior of the cuprates. We can also explain the negative shift conundrum and the long-standing orbital shift discrepancy for NMR in the cuprates.\end{abstract}
\maketitle
\labelformat{paragraph}{#1}

\subsection{Introduction}
Nuclear spins are powerful quantum sensors of their local electronic environment, so that the versatile methods of nuclear magnetic resonance (NMR) can be decisive for theories of condensed matter systems. However, deciphering the nuclear response is usually not a straightforward task if microscopic theory is missing, as is the case for high-temperature superconducting cuprates. Nevertheless, NMR contributed vital information for the understanding of these materials, early, e.g., concerning singlet pairing and the pseudogap \cite{Slichter2007, Walstedt2008}. 
Through magnetic shift and nuclear relaxation, NMR can sense the field-induced electronic moments and local fluctuating fields, respectively, both related to the electronic susceptibilities. 
In addition, the electric quadrupole interaction, e.g., of Cu and O nuclei in the ubiquitous \ce{CuO2} plane, allows for the determination of the local charge \cite{Pennington1989, Zheng1995, Haase2004}.

For useful conclusions, however, the hyperfine interactions have to be known. For the electric interaction a convincing understanding could be achieved only recently \cite{Jurkutat2014}, which led to, e.g., the discovery of the correlation between the sharing of charge between planar Cu and O, and the maximum \tc \cite{Rybicki2016}, as well as the measurement of charge ordering in the unit cell \cite{Reichardt2018}. The fact that the charges at planar Cu and O are shared quite differently between the families that have different maximum \tc suggests that one might expect fundamental differences between different cuprate families in terms of magnetic shift and relaxation, as well.

The \textit{magnetic} hyperfine scenario was established rather early \cite{Takigawa1989,Takigawa1991}, predominantly on the \ybco family of materials. Here, the consequences of the apparent \mbox{3$d(x^2-y^2)$} hole of \ce{Cu^{2+}} were investigated, and indeed, the quadrupole splitting of Cu was  found to be in qualitative agreement with such a hole \cite{Pennington1989}. However, and surprisingly, a related negative spin shift that must arise from such a hole is not observed \cite{Pennington1989,Takigawa1989}. Rather, the total shift was found to be positive \cite{Takigawa1989}. Moreover, at lower doping the shift is temperature (\temp) dependent, even above the superconducting transition temperature (\tc), which marked the discovery of a spin gap above \tc \cite{Alloul1989}. However, this shift was not \temp-dependent for Cu when the magnetic field is perpendicular to the CuO$_2$ plane. 
\begin{figure}[H]
\centering
\includegraphics[width=0.22 \textwidth ]{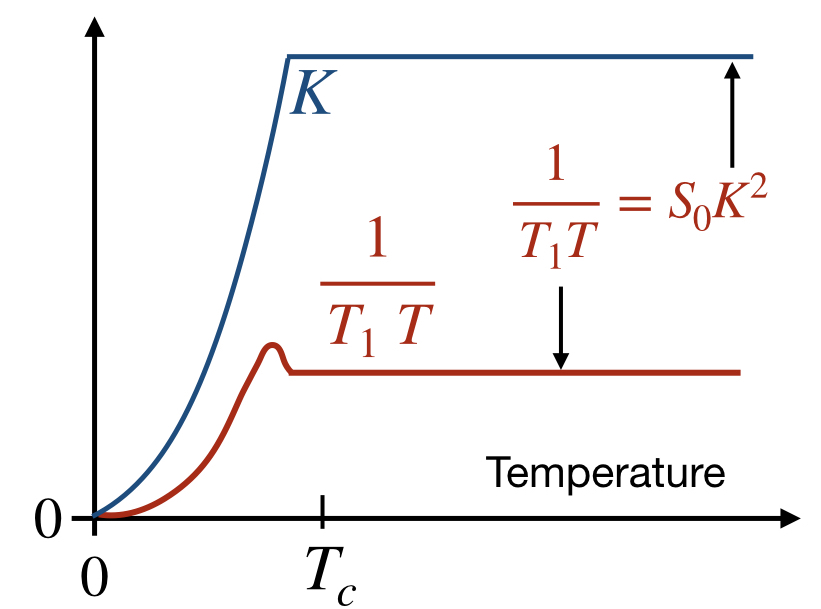}
\caption{Fermi liquid shift and relaxation: the spin shift $K$ is temperature independent above the critical temperature for superconductivity ($T_{\rm c}$) and vanishes below it for spin singlet pairing. The nuclear relaxation rate ($1/T_1$) is proportional to the temperature ($T$) so that $1/T_1T$ is temperature independent above $T_c$, and the Korringa relation holds roughly, where $S_0 = {\gamma_n^2}/{\gamma_e^2}{4\pi k_B}/{\hbar}$ contains only fundamental constants, e.g., the nuclear ($n$) and electronic ($e$) gyromagnetic ratios. $1/T_1T$ vanishes below $T_c$ for singlet pairing. In early analyses of the cuprates it was found that $1/T_1T$ is much larger than what was expected from $K$, and special spin fluctuations were invoked to account for the discrepancy.}\label{fig:float}
\end{figure}
This mysterious behavior was interpreted as an accidental cancellation of the spin shift from a very anisotropic sum of hyperfine coefficients $A_\alpha + 4B'$, where $A_{\alpha}$ is anisotropic due to the $3d(x^2-y^2)$ orbital, and $B'$ an isotropic transferred coefficient from the neighboring four Cu atoms in a single band scenario ($B'$ in order to distinguish it from $B$ as introduced later). With this explanation of the shifts a single electronic spin component could be salvaged, and was supported by measurements on two materials \cite{Takigawa1991,Bankay1994}. One problem with the understanding of the shifts is the separation of orbital and spin shift contributions, and one adopts the following chain of arguments. Orbital shift is not temperature dependent, and since a temperature dependent component is observed, it is taken as the spin shift term. Furthermore, since there is singlet pairing, the spin shift should nearly disappear at low temperatures for all directions of the field. The thus deduced orbital shifts fit the single ion estimates for Cu orbital shift \cite{Pennington1989}, but were not expected to hold for the realistic chemistry of the CuO$_2$ plane \cite{Pennington1989}.

Nuclear relaxation data, also mostly on the \ybco family of materials, were hampered by the assignment of Cu sites in the plain and chain. The Y nucleus, situated between the two CuO$_2$ planes, showed Fermi liquid relaxation, as well as one Cu site. Surprisingly, it turned out later that the chain site, rather than the planar Cu site was more Fermi liquid-like. More importantly, it appeared that the Korringa relation did not hold and that there must be an about 10-fold increase of relaxation \cite{Walstedt1988}. With $A_\parallel + 4B' = 0$ in one direction, antiferromagnetic fluctuations in a single band scenario would turn this term into a large pre-factor $|A_\parallel - 4B' | \gg 0$ (correlations between neighboring spins are negative), and one could explain the data. With most available measurements for \cpara (aligned powders and NQR) and material and sample dependent rates, many approaches were developed to understand shift and relaxation.

Over the last ten years, it was shown with a set of experiments on different materials that the adopted single spin component view does not hold, rather, two coupled spin components appear to be necessary to explain the NMR shifts \cite{Haase2009b,Meissner2011, Haase2012, Rybicki2015}. 
Finally, a simple literature survey of all Cu NMR shifts uncovered significant differences between the cuprates, which point directly to a new shift phenomenology, at odds with the hitherto used hyperfine scenario, and that cannot be understood with a single spin component \cite{Haase2017}.

These findings raised the interest in nuclear relaxation, e.g. the apparently large isotropic hyperfine coefficient, and we began gathering the Cu relaxation data. 
Here, we show that the Korringa relation does nevertheless hold for some cuprates, i.e. those with the highest doping levels. And since the nuclear relaxation is rather similar for all cuprates (only its anisotropy changes among the systems) a suppression of the spin shifts for certain cuprates is behind the failure of the Korringa relation for these systems, not an enhancement of nuclear relaxation.  In particular, there cannot be substantial spin fluctuations, except for rather low doping levels where there are no NMR data, and for one outlier system, \lsco, that we find has an additional relaxation mechanism. We propose that a negative coupling between two electronic spin components already found in 2009 \cite{Haase2009} is behind the suppression of the shifts, while hardly affecting a universal Fermi liquid-like relaxation. 

\subsection{Observations from shifts and relaxation}
We begin with an overview of results from literature shift and relaxation analyses. For a full review of literature shifts see \cite{Haase2017}, and after the first submission of this manuscript a more comprehensive review of the relaxation was prepared, as well \cite{Jurkutat2019}. 

Throughout the manuscript, quantities measured with the magnetic field ($B_0$) parallel to the crystal $c$-axis (\cpara), are labeled like $\hat{K}_\parallel$ (total magnetic shift), $K_\parallel$ (spin shift), or $1/T_{1\parallel}$ (also the relaxation in NQR measurements carries the same label since it is dominated by crystallites for which the nuclear quantization axis is parallel to $c$ due to quadrupole interaction). Measurements with the magnetic field in the CuO$_2$ plane (\cperp) are labelled like $\hat{K}_\perp$, $K_\perp$, or $1/T_{1\perp}$; note that measurements along special in-plane axes are very rare since $c$-axis aligned powders are most easily measured and twinning can be a problem even for single crystals.\par\medskip
\begin{figure}[!]
\centering
\includegraphics[scale=0.22]{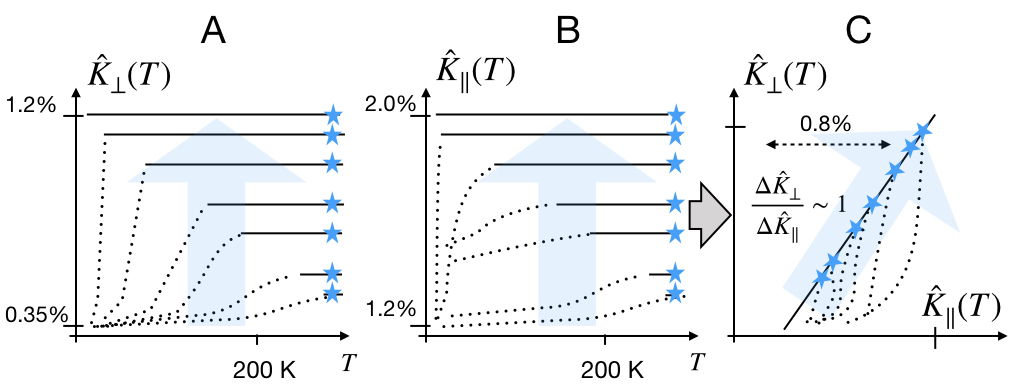}
 \caption{Sketch of cuprate \emph{total} magnetic shifts, A, $\hat{K}_\perp(T)$, and B, $\hat{K}_\parallel(T)$. Increasing the doping (blue arrow) increases the temperature independent shifts that extend to lower temperatures. Shifts become temperatures dependent at lower $T$ and head for a common value for $\hat{K}_\perp$, but not for $\hat{K}_\parallel$. C, a useful overview of the shift is obtained by plotting $\hat{K}_\perp(T)$ vs. $\hat{K}_\parallel(T)$. The high temperature shifts fall on a more or less straight line with slope ${\sim1}$, i.e., an 'isotropic shift line' is created by changes in doping, which demands a dominant isotropic hyperfine coefficient. As the shifts become temperature dependent they depart from the isotropic shift line with special slopes, but stay in the lower right triangle.
}
\label{fig:fig2}
\end{figure}

\subsubsection{Magnetic shifts}
As mentioned in Introduction,  early experiments showed that $\hat{K}_\parallel$ is \temp-independent, while $\hat{K}_\perp(T)$ appears to have a Fermi liquid-like spin component near or above optimal doping in the sense that one observes a decrease as \temp is lowered. This decrease is more abrupt when it occurs at \tc, but it can also start far above \tc (which is the assumed pseudogap behavior).  So it was argued that $\hat{K}_\parallel $ contains no spin shift (${K}_\parallel = 0$), only orbital shift contributions ($K_{\rm L \parallel}\approx \hat{K}_\parallel$).  The spin shift was only extracted from $\hat{K}_\perp$ by defining as spin shift $K = \hat{K}(T \rightarrow 0)$. Sine the spin response should be isotropic in the cuprates, this anisotropic behavior was explained with anisotropic hyperfine coefficients, i.e. $A_\parallel + 4B'=0, A_\perp + 4B' \neq 0$ (in the original literature it was $B$, not $B'$) \cite{Walstedt2008}. 

Later, materials were investigated that also showed a significant temperature dependence for $K_\parallel(T)$ (for references see \cite{Haase2017}), however, their shift anisotropy was found to be temperature dependent \cite{Haase2012,Rybicki2015}, which is not expected in a single spin component scenario. Furthermore, an explanation within the old scenario would require rather different hyperfine coefficients (up to about 30\%), which appears to be unrealistic given the unique CuO$_2$ plane.

\begin{figure}
\centering
\includegraphics[scale=0.25]{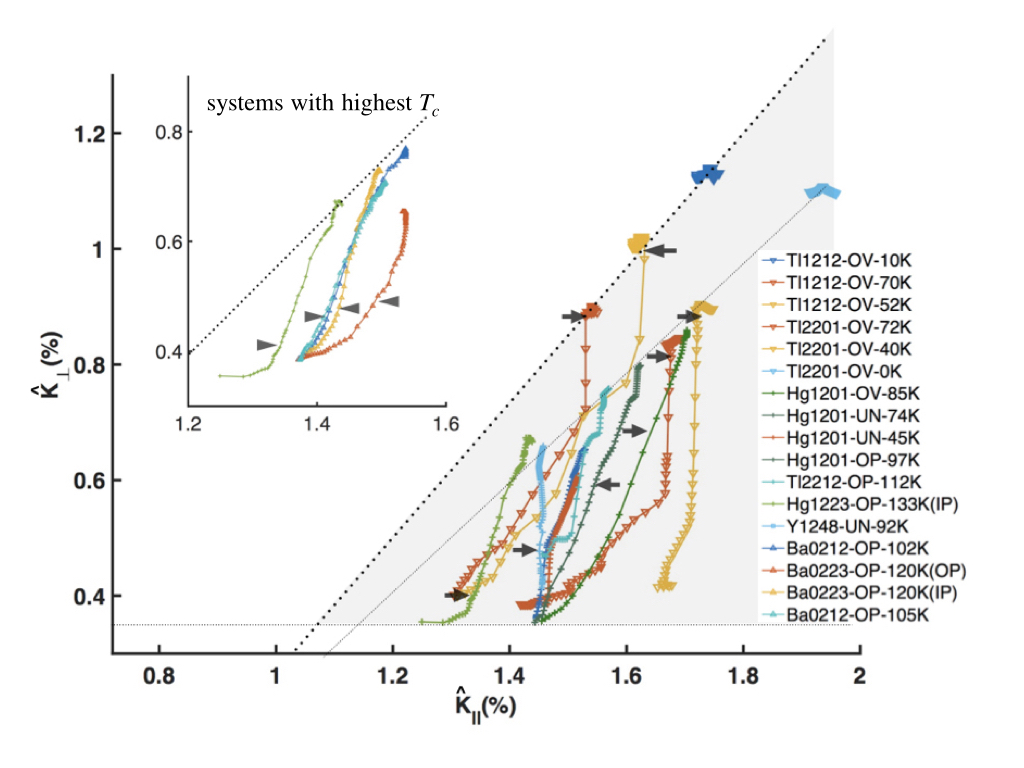}
 \caption{Examples of $^{63}$Cu NMR shifts from Ref.~\cite{Haase2017}. $\hat{K}_\perp (T_j)$ is plotted against $\hat{K}_\parallel (T_j)$ (the \emph{hat} denotes the total magnetic shifts, including orbital shifts). The plot origin reflects first-principle calculations of the orbital shifts of $K_{\parallel \rm L} = 0.72\%, K_{\perp \rm L} = 0.3\%$ \cite{Renold2003}. Nearly isotropic shift lines are indicated by dashed lines. The maximum high temperature shift increases with doping ($x$). The arrows indicate $T_{\rm c}$ (OD: overdoped; OP: optimally doped; UN: underdoped materials, cf. Appendix). Inset: Materials with the highest \tc depart from the isotropic shift line at the NMR pseudogap temperature far above \tc, unlike strongly overdoped systems in the main panel, for which \tc determines the departure point. }\label{fig:jh7}
\end{figure}

A convenient and useful overview of both shifts can be obtained by plotting $\hat{K}_\perp(T)$ vs. $\hat{K}_\parallel(T)$ \cite{Haase2017}. This is illustrated in Fig.~\ref{fig:fig2} that should help understand the real data plotted in the same way in Fig.~\ref{fig:jh7}. Note that the total shifts $\hat{K}$ are plotted in order to avoid a biased analysis by subtracting unknown orbital shifts. 

Key features of such a plot are the following, cf.~\cite{Haase2017}. There is a common low temperature shift  for \cperp, $\hat{K}_\perp ({T\rightarrow0}) \approx 0.35\%$. It agrees reasonably well with first-principle calculations of the orbital shift that give 0.30\% \cite{Renold2003}. Interestingly, $\hat{K}_\parallel ({T\rightarrow0})$ can be very different for different cuprates.
Different families have slightly different \emph{isotropic shift lines} (defined by changing doping at high temperatures). New shift reference points are generated where $\hat{K}_\perp ({T\rightarrow 0}) \approx 0.35\%$ intersects isotropic shift lines. These points could define $K_{\rm L\parallel}$, but they are still in strong disagreement with the calculated orbital shift of 0.72\% \cite{Renold2003} (for the figure origin we assumed $K_{\rm L\parallel,\perp}$ as found from first-principle calculations).\par
Differently from changes due to doping, as the shifts change as a function of temperature their anisotropy changes, i.e., the shifts depart from the \emph{isotropic shift line} in Fig.~\ref{fig:jh7}. However, the slopes with respect to temperature, $\delta_T \hat{K}_\perp/\delta_T \hat{K}_\parallel$, appear to be constant in certain ranges of \temp, which causes the characteristic linear regions in that figure. Characteristic slopes as a function of \temp are: (1) $\delta_T K_\perp/\delta_T K_\parallel \approx 1$ (same slope as the isotropic shift lines, but here as function of \temp); (2) a rather steep slope $\delta_T K_\perp/\delta_T K_\parallel \geq 10$, and (3) $\delta_T K_\perp/\delta_T K_\parallel \approx 5/2$. This has been discussed in more detail previously \cite{Haase2017}. \label{par:a5} \par\medskip

Perhaps the most surprising fact concerns the nearly isotropic change in shift as function of doping, as this points to a large isotropic hyperfine coefficient that has not been discussed so far. In addition, since a variation in temperature can lead to different slopes, one must conclude that different spin components are at play that couple to the nucleus. We do not see a possibility to account for the shift scenario with a single temperature and/or doping dependent spin component \cite{Haase2017}. These are similar conclusions to those deduced with very different shift experiments \cite{Haase2009,Meissner2011,Haase2012,Rybicki2015}.\par\medskip

\subsubsection{Nuclear relaxation}
Clearly, given the different phenomenology that appears from viewing all the cuprate shifts, one has to take an unbiased look at relaxation data as well. After the first presentation of our short relaxation summary here, we prepared a more comprehensive account that is available now, as well \cite{Jurkutat2019}.

\begin{figure}
\centering
\includegraphics[scale=0.22]{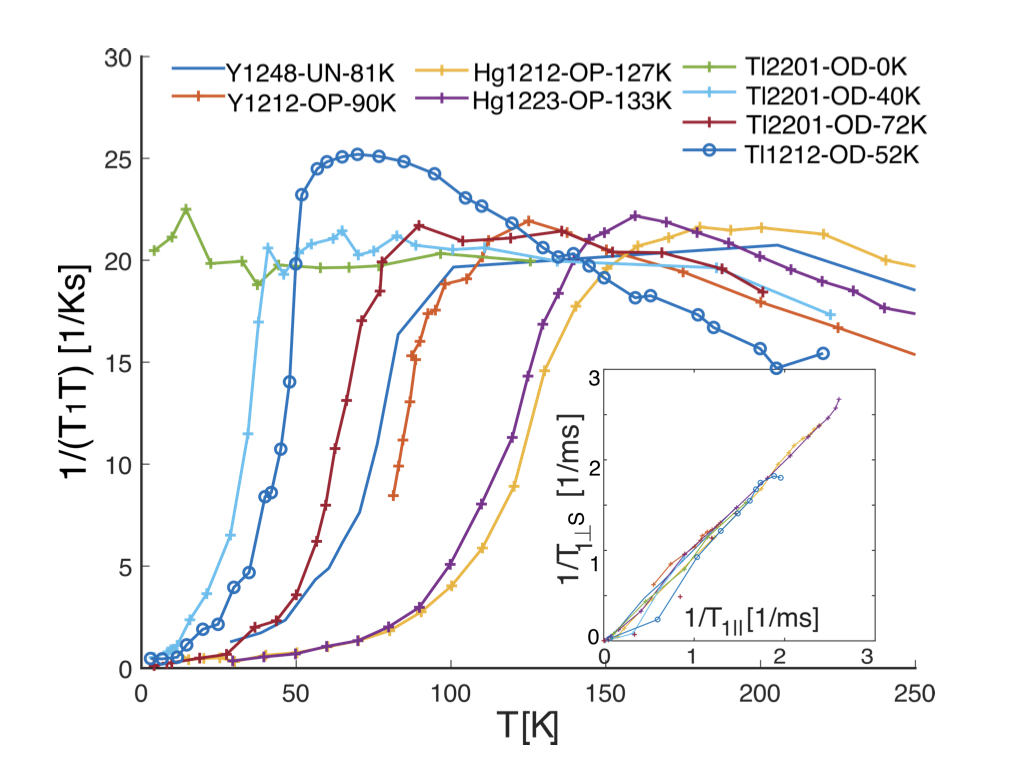}
 \caption{$^{63}$Cu NMR relaxation rates for various materials. Main panel: $1/T_{1\perp}T $ is very similar above \tc for all systems (even those that do not superconduct) and Fermi liquid-like, i.e., $1/T_{1\perp}T = const.$ above \tc and disappears below \tc from singlet pairing. From Korringa's relation and a Knight shift of 0.8\% (cf.~Fig.~\ref{fig:jh7}) one finds $1/(T_{1}T) \approx \SI{20}{/Ks}$. Inset: $(1/T_{1\perp})/(s\cdot (1/T_{1\parallel})$ of the same materials with \temp independent proportionality factor $s = 3.3, 3.1, 2.0, 1.9, 1.0, 1.5, 1.5, 1.7$ for the systems according to their appearance in the legend at the top. The rates are proportional to each other above \tc where the orientational dependence of the field is expected to be irrelevant for the fluid; even below \tc only a couple of strongly overdoped materials show an slight deviation.}
\label{fig:ma1} 
\end{figure}

The few outstanding observations from viewing the relaxation data are the following. First, the relaxation rate $1/T_{1\perp}$ measured for \cperp is rather similar for all cuprates, above \tc. In particular below about \SI{200}{K}, the most overdoped system that is not superconducting has a similar Fermi liquid-like dependence as an underdoped cuprate (there are not enough data to conclude on strongly underdoped systems). This is seen in Fig.~\ref{fig:ma1} where we plot typical examples (for more data see \cite{Jurkutat2019}). Second, $1/T_{1\parallel}$ behaves differently, but as we show in the inset of Fig.~\ref{fig:ma1} both rates are nearly proportional to each other, above \emph{and} below \tc. Thus, it is only the relaxation anisotropy that changes among the systems and with doping, from $(1/T_{1\perp})/(1/T_{1\parallel}) \approx 1 {\rm \;to\;} 3.4$.
Since there was a clear emphasis on $1/T_{1\parallel}$ measurements and since some systems were investigated only later, this behavior was not discovered (however, Walstedt et al. noted the anisotropy for \ybcoF \cite{Walstedt1988}).

In particular, the relaxation just above \tc is very similar for all (conducting) cuprates in terms of $1/T_{1\perp}T$, i.e., it is material independent and it does not change very much across the phase diagram. Just above \tc we have $1/(T_{1\perp}T) (T \gtrsim T_{\rm c}) \approx 17 \text{\;\;to\;\;} \SI{25}{/Ks}$, which gives a shift of about 0.8\% from the Korringa relation. Note that this is the maximum shift observed in Fig.~\ref{fig:jh7}. \par

Third, if one includes the differences in anisotropy among different families of materials, the relaxation rates below \tc are very similar, as well (see below). \par\medskip

\subsubsection{Shifts and relaxation}
The fact that the nuclear spins are coupled to an electronic thermal bath with relaxation rates that are nearly independent on material and doping (in particular near \tc) points to a very robust property. This special relaxation rate with $1/T_{1\perp}T \approx \SI{20}{/Ks}$ is already present at the highest doping levels for systems that must be rather close to a Fermi liquid. It appears to be out of question, then, that these excitations (this liquid) is present in all materials and dominates relaxation above \tc (at higher temperatures the rate lags somewhat behind, and $1/(T_1T)$ falls off in a characteristic way for more or less all the systems). This conclusion is not weakened by a doping or material dependent relaxation for the other direction of the field (\cpara) since both rates are proportional to each other. It rather points to an anisotropic coupling of the nuclear spins to a unique fluid, which can depend on doping and material. Interestingly, the anisotropy takes on only special values (reminding one of selection rules, rather than a crossover). 

\begin{figure}
\centering
\includegraphics[scale=0.22]{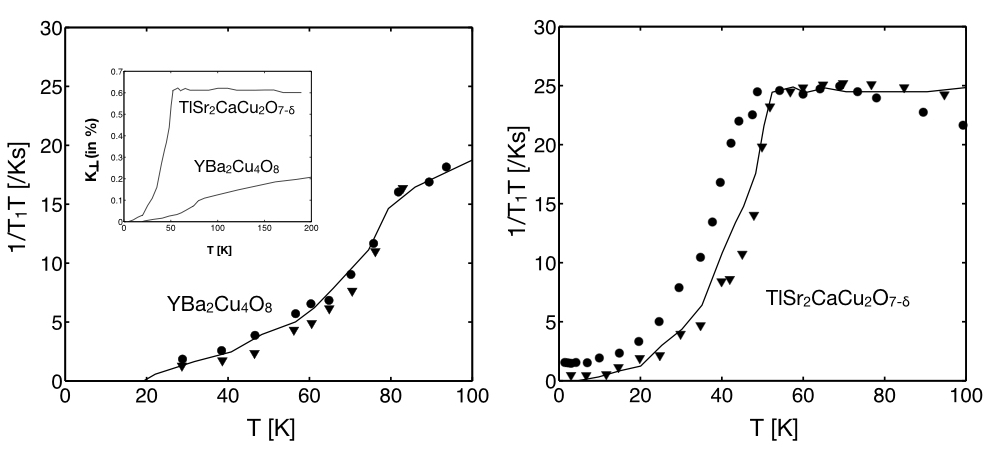}
 \caption{Comparison of $K_\perp$ and $1/(T_{1\perp,\parallel}T)$ below \tc for an underdoped and overdoped material. The relaxation rates for \cpara (circles) and \cperp (diamonds) are plotted as a function of temperature for the \ce{YBa2Cu4O8} (\tc $\approx \SI{81}{K}$) and \ce{TlSr2CaCu2O$_{7-\delta}$} (\tc $\approx \SI{52}{K}$); note that $1/T_{1\parallel}$ is multiplied by the proportionality constants above \tc given in Fig.~\ref{fig:ma1} (1.5 for \ce{TlSr2CaCu2O$_{7-\delta}$} and 3.3 \ce{YBa2Cu4O8}). The original shifts shown in the inset are scaled in the main panel by a ratio of 15/4 to fit the relaxation curves.}
\label{fig:ws}
\end{figure}

Below \tc, the relaxation rates for both directions of the field drop rapidly, probably from spin singlet pairing. Both rates are nearly proportional to each other. Interestingly, the perpendicular shift, $K_\perp(T)$, that approaches a common value for all cuprates appears to be nearly proportional to $1/(T_{1\parallel,\perp}T)$. However, the proportionality factor depends on the maximum shift for that material, as can be seen in Fig.~\ref{fig:ws}. This tells us that the shifts for materials that are located in the lower left part of Fig.~\ref{fig:jh7} are suppressed compared to those in the upper right section of the plot that is reached for certain families and at high doping levels.

It is not quite obvious whether it is just the doping level that would bring all cuprates in the upper right corner of Fig.~\ref{fig:jh7}. We also know that there is a correlation between the sharing of the charge in the CuO$_2$ plane and the maximum \tc, which is not apparent in terms of the total doping (the sum of the planar Cu and O holes), which proves that doping is not \emph{not} the key parameter for all properties. Therefore, we introduce a parameter $\zeta$ (that clearly depends on doping) to be the cause of the changes of the uniform response, in addition to \temp, i.e., we write $\chi_0(\zeta,T)$.

The question arises how one can reconcile a robust and material independent relaxation with a suppressed high-temperature shift. Of course, the uniform response, $\chi_0(T)$, can be very different from the wavevector ($q$)  dependent imaginary part of the susceptibility, $\chi''(q,T)$. For example, a sinusoidal modulation of the spatial spin response will reduce the uniform response of the system, but can leave the local fluctuations that set relaxation unchanged. 

We do know from experiments on a number of different cuprates \cite{Haase2009,Meissner2011,Haase2012,Rybicki2015} that a single electronic spin component cannot explain the shift data, rather at least two components appear to be necessary, and couple to the nucleus with different hyperfine coupling constants to the electronic excitations. Therefore, a simple uniform $\chi_0(\zeta,T)$ is not sufficient to explain the data, that is why we propose a simple two-component model.

\subsection{Simple two-component description}
In the most simple two-component model, the nuclear spin couples to two electronic spin components with the susceptibilities $\chi_{\rm A}$ and $\chi_{\rm B}$. These spin components will then have different \temp dependences in general. We write,
\begin{equation}
K_{\parallel, \perp} (\zeta,T) = B_{\parallel, \perp} \cdot \chi_{\rm B}(\zeta,T) + A_{\parallel, \perp} \cdot \chi_{\rm A}(\zeta,T)
\label{eq:shift00}\end{equation}
With other words, the magnetic field ($B_0$) induces the two electronic spin components $\erww{S_{\rm A}}$ and $\erww{S_{\rm B}}$ ($\gamma_e \hbar \erww{S_{\rm j}} = \chi_{\rm j} B_0$), which are not proportional to each other as a function of temperature. 
The Cu nucleus feels changes in the local field through the corresponding hyperfine coefficients $B_{\parallel, \perp}$ and $A_{\parallel, \perp}$. 

Now, we denote with $B$ the apparently isotropic hyperfine coefficient that arises as a function of $\zeta$ in Fig.~\ref{fig:jh7} (we do not invoke the factor of 4, as opposed to the old literature). Then, there must also be an anisotropic local field contribution. In a minimalistic model, we seek this component in terms of the partially unfilled $3d(x^2-y^2)$ orbital. As in the early literature we denote this coefficient with $A_{\parallel, \perp}$. It is also known from reliable estimates, as well as experiment \cite{Pennington1989} that
\begin{equation}\begin{split}
|A_\parallel | \gtrsim 6 |A_\perp| \text{  and } A_\parallel =-|A_\parallel |,
\end{split}\end{equation}
i.e., the anisotropic hyperfine coefficient is negative and must lead to a negative shift for a positive spin moment.
We will neglect the smaller $|A_\perp|$, and we have with \eqref{eq:shift00},
\begin{equation}\begin{split}
&K_{\perp}(T) = B\cdot \erww{S_{\rm B}}(T),\\
&K_\parallel(T) = B \cdot \erww{S_{\rm B}}(T) + A\cdot \erww{S_{\rm A}}(T),
\end{split}\end{equation}
where $A \equiv A_\parallel$. Again, we seek to explain the NMR shift with these two equations that follow from the experimental observation that $\delta_T K_\perp(T)$ is \textit{not} proportional to $\delta_T K_\parallel(T)$, and the fact that we need two different hyperfine coefficients, plus the assumption that one coefficient is isotropic and the second is related to the partially filled $3d(x^2-y^2)$ orbital.

If two spin components are present we must allow for a coupling between them \cite{Haase2009}. Thus, each spin component is the sum of two terms, and we use the simplified notation,
\begin{equation}
\erww{S_{\rm B}} \equiv b + c, \;\;\; \erww{S_{\rm A}} \equiv a + c,
\label{eq:comp}\end{equation}
where the spin components are denoted by $a(\zeta,T), b(\zeta,T)$ and the coupling term by $c(\zeta,T)$.

That is, we have to analyze the shifts in Fig.~\ref{fig:jh7} with the following two equations,
\begin{equation}\begin{split}\label{eq:shift01}
K_{\perp} (\zeta, T) &= B \big[b(\zeta, T) + c(\zeta, T)\big]\\
K_\parallel (\zeta, T) &=  A \big[a(\zeta, T) + c(\zeta, T)\big] + B \big[b(\zeta, T) + c(\zeta, T)\big],
\end{split}\end{equation}
where $T$ is the temperature, and $\zeta$ takes care of the material related property.\par\medskip

We now investigate some consequences in this simple picture, and we begin with the low temperature shifts for \cperp. We remember that, $\hat{K}_\perp(T\rightarrow 0) \approx 0.35\%$ is rather similar for all cuprates, and second, it agrees reasonable well with first principle calculations that predict 0.30\% \cite{Renold2003}. Therefore, we make the fundamental assumption that $\hat{K}_\perp (T = 0) = K_{\rm L\perp}$ is the orbital shift for this orientation of the field, i.e., the spin shift is zero (singlet pairing). We then conclude with \eqref{eq:shift01},
\begin{equation}\label{eq:ass01}\begin{split}
b(\zeta, T\rightarrow 0) + c(\zeta, T\rightarrow 0) \approx 0.
\end{split}\end{equation}
Note that we only know the sum ($b+c$) vanishes at low temperature, not each component separately.\par

Next, we address the isotropic shift lines that appear at high temperatures in Fig.~\ref{fig:jh7}. They demand that the changes in the shifts induced by $\zeta$, i.e. $\delta_\zeta K_\alpha$, are nearly proportional to each other, i.e.,
\begin{equation}\label{eq:shift03}
\delta_\zeta K_\perp \approx \delta_\zeta K_\parallel,
\end{equation}
and it follows with \eqref{eq:shift01},
\begin{equation}\label{eq:shift03b}\begin{split}
&\delta_\zeta (a+c) \approx 0.
\end{split}\end{equation}
That means, the material related shift variations at high \temp are given by $\delta_\zeta K_{\perp,\parallel} = B \delta_\zeta (b+ c)$, i.e., for both orientations of the field. \par\medskip

With \eqref{eq:ass01}, we assumed the orbital shift for \cperp to be given by $K_{\rm L\perp} \approx 0.35 \%$ (as in the old model for the hyperfine scenario). Since the orbital shift anisotropy of 2.4 calculated from first principles is a rather reliable number \cite{Renold2003}, we conclude that $K_{\rm L\parallel} \approx 0.84\%$ is a  reliable orbital shift value for \cpara, as well. This is very different from the old scenario where the orbital shift for \cpara was defined by the \ybco{} low \temp shift.

In our two-component analysis at the (virtual) intersection of an isotropic shift line with $K_{\rm L\perp} \approx 0.35\%$, which defines $\zeta \equiv \zeta_\Lambda$, we have,
\begin{equation}\label{eq:cross}
K_\parallel (\zeta_\Lambda, T_h) = A \big[a(\zeta,T_h) + c(\zeta,T_h)\big],
\end{equation}
where $T_h$ was introduced to denote a sufficiently high \temp, i.e. $T \gg T_{\rm c}$.
This is the material independent offset of the isotropic shift lines in Fig.~\ref{fig:jh7}. Near the intersection $\zeta_\Lambda$ we have with \eqref{eq:shift01} that $K_\perp (\zeta_\Lambda, T) = B \big[ b(\zeta_\Lambda,T) + c(\zeta_\Lambda, T)\big]$, where $K_\perp (T)$ is very small even at high \temp. Thus, $c(\zeta_\Lambda) = - b(\zeta_\Lambda)$ holds to a good approximation for all \temp. We thus have in addition to \eqref{eq:cross},
\begin{equation}\label{eq:cross2}\begin{split}
K_\parallel (\zeta_\Lambda, T_h) &= A \big[a(\zeta_\Lambda,T_h) + c(\zeta_\Lambda)\big]\\
K_\parallel (\zeta_\Lambda, T_h) &= A \big[a(\zeta_\Lambda,T_h) - b(\zeta_\Lambda)\big].
\end{split}\end{equation}
With $K_{\rm L\parallel} = 0.84\%$ we have,
\begin{equation}\label{eq:cross3}
K_\parallel (\zeta_\Lambda, T_h) = 0.21\%.
\end{equation}
Clearly, there could be  differences between the materials in terms of $\big[a(\zeta)+c(\zeta)\big]$, but also the orbital shifts could vary slightly. However, it must be the negative coupling term $c(\zeta,T)$ that is responsible for the positive offset in the spin shifts of the cuprates for \cpara.
With other words, there is an effective negative spin in the $3d(x^2-y^2)$ orbital, while component $a$ itself is positive, and at high \temp $(a+c)$ does not change with $\zeta$. 
\begin{figure}
\centering
\includegraphics[scale=0.24]{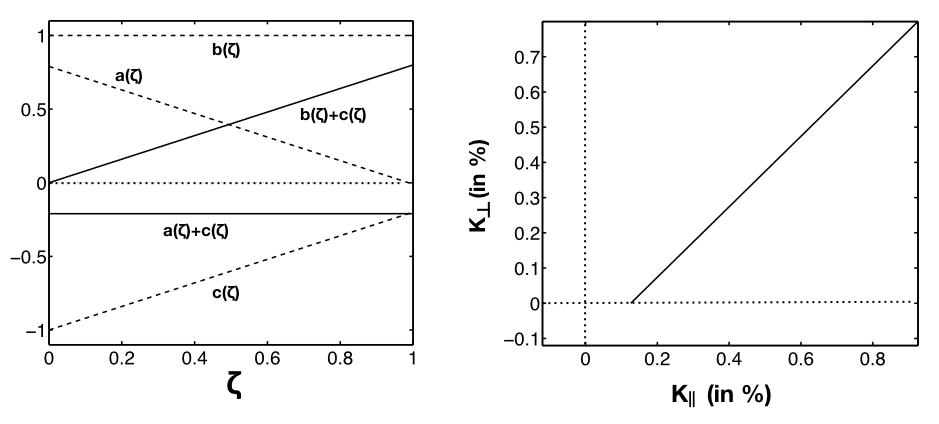}
 \caption{Left, possible decomposition of the high-\temp shifts as function of the material related parameter $\zeta$ and the spin components $a, b$ and the coupling $c$, cf.~\eqref{eq:shift01}. Right, resulting shift-shift plot according to \eqref{eq:shift01} for $B=1$ and $A=-3/5 B$.}
\label{fig:components}
\end{figure}

We note that the maximum shift variation above the intersection defined by $\zeta_\Lambda$ is about $0.8\%$, and we conclude that
\begin{equation}\label{eq:bterm}
B\cdot \big(b(\zeta_{\rm max},T_h)-b(\zeta_{\Lambda},T_h)\big) \approx 0.8 \%.
\end{equation}
Roughly, there is a factor of 4 between $K_\parallel (\zeta_\Lambda,T_h)$ and the maximum $\zeta$-related shift change.
Note that an isotropic shift of about 0.8\% is in agreement with the observed universal relaxation rate just above \tc, i.e., it follows from Korringa's law for a simple Fermi liquid. \par\medskip

Based on the above discussion we present in Fig.~\ref{fig:components} a possible decomposition of the high-\temp shifts, and the ensuing shift-shift plot, inspired by a large $b$ term from a robust Fermi liquid-like fluid, a negative coupling $c$ that tries to align positive spin components $a$ and $b$ antiferromagnetically.\par\medskip

Now, we turn to the temperature dependence of the shifts. The fact that basically all shift data lie below the isotropic shift lines in Fig.~\ref{fig:jh7} tells us that as the shifts depart from the isotropic shift lines with $\delta_T K_\perp \leq \delta_T K_\parallel$. It follows with \eqref{eq:shift01},
\begin{equation} \label{eq:triangle}\begin{split}
\delta_T [A (a+c)] & \gtrsim 0\\
\delta_T (a + c) & \lesssim 0,
\end{split}\end{equation}
since $A$ is negative. This says that by lowering the temperature, $A\big(a(T)+c(T)\big)$ becomes more positive so that $K_\parallel$ stays to the right of the isotropic shift lines in Fig.~\ref{fig:jh7}. The nearly equal sign refers to points very near the isotropic shift line. \par\medskip

In Fig.~\ref{fig:jh7} we pointed to certain slopes in the low temperature behavior of the shift anisotropies (for a more detailed discussion see \cite{Haase2017}).\par
First, we have $\delta_T K_\perp/\delta_T K_\parallel \approx 1$, similar to the isotropic shift lines, but now as a function of \temp. We conclude $\delta_T (a + c) \approx 0$. This slope is observed, in particular, for overdoped systems where, after an initial steep drop of $K_\perp$ at \tc, the system holds $(a+c)=const.$ as \temp drops further, cf.~Fig.~\ref{fig:jh7}. We do know that $(b+c)$ varies in this range of \temp since $K_\perp$ changes. 

Second, we find in Fig.~\ref{fig:jh7} the steep slope, i.e. $\delta_T K_\perp/\delta_T K_\parallel \gtrsim 10$. It can be found for the strongly doped systems at \tc for a given range of \temp, but also for other materials, e.g., \ybcoE in the whole range of \temp. This includes the variation in the NMR pseudogap region, but not for all materials. For example, \hgryb takes on the slope of $\approx 5/2$ as it departs from the isotropic shift line at \tc, or in the pseudogap region. With $\delta_T K_\parallel \approx 0$ we conclude that 
\begin{equation}\label{eq:steep}
B\; \delta_T (b+c) \approx -A\; \delta_T (a+c).
\end{equation}
If only $c$ became \temp dependent, $A = - B$ would follow, the known argument in the old literature (our definition of $B$ is that of $4B'$ in those papers).

Third, we observe a typical slope of $\delta_T K_\perp/\delta_T K_\parallel \approx 5/2$. This leads to the equation,
\begin{equation}\label{eq:5o2}
B\; \delta_T (b+c) \approx -\frac{5}{3}A\; \delta_T (a+c).
\end{equation}
For example, if we assume that only $c$ changes as a function of \temp for those slopes, we conclude that $B \approx -5/3 A$. This is perhaps a reasonable conclusion, and the \temp dependent NMR pseudogap feature is caused by a \temp dependent $c(\zeta)$. Then, in order to generate, e.g., the steep slope, we find $3 \delta_T b = 2 \delta_T a-\delta_T c$. \par\medskip

The behavior of the shifts at low temperatures is perhaps more complicated. One must also be aware of the fact that the measurements were not pursued with the appropriate rigor since such behavior was not suspected. In addition, the penetration depth of the r.f. decreases rapidly and signal-to-noise can become a limiting factor, certainly for single crystals. Perhaps, $K_{\rm L\perp} = 0.35\%$ is somewhat higher than the calculated 0.30\%. We cannot be sure that all $\hat{K}_\parallel(\zeta) (T=0)$ in Fig.~\ref{fig:jh7} are the true low-\temp shifts for this orientation. If so, we clearly need negative spin $a+c$, i.e., $K_\parallel = -|A| (b-a)$ if $c=-b$. For example, the single layer \hgryb \cite{Rybicki2016} has a $T=0$ shift of $K_\parallel = +0.6 \%$, and we conclude that $A(a+c)$ increased 3-fold compared to the $\zeta_\Lambda$ value of 0.21 \%.\par\medskip

How can one reconcile the variations of shift and relaxation? First, we focus on the largest $\zeta$ materials, which show Fermi liquid-like behavior with an isotropic $1/T_1T$ of about \SI{20}{/Ks}. This value follows from the Korringa relation for a simple Fermi liquid. Note that $(1/T_{1\perp})/(1/T_{1\parallel}) \approx 1$ is expected for relaxation dominated by fluctuations through $B$. Thus, the largest-$\zeta$ systems are easily understood.

As $\zeta$ decreases, the shifts decrease isotropically with decreasing $\zeta$, but remain \tind above \tc. The relaxation is strictly proportional to \temp and even remains very similar, except that the anisotropy changes to $(1/T_{1\perp})/(1/T_{1\parallel}) = 1.5$. By decreasing the temperature, \tc is encountered and the shifts suddenly drop. First, $K_\perp$ begins to change, the initial steep drop in Fig.~\ref{fig:jh7}. It is followed by a nearly proportional decrease of both shifts along isotropic shift lines, now as a function of \temp. The initial drop can be rather large, followed by a short isotropic shift line to reach $K_{\rm L\perp} (0) \approx 0.35\%$. Systems with a small initial drop have a longer isotropic shift line since it ends at $K_{\rm L\perp}$. Consequently, in the latter case a smaller shift ($K_\parallel (T=0)$) remains at the lowest \temp. While the changes in the shifts are more complex, both relaxation rates drop almost proportionally to $K_\perp$ below \tc (and they are nearly proportional to each other). This is expected for singlet pairing, here as vanishing of $b+c$. $1/(T_{1\perp,\parallel}T)$ is nearly proportional to $K_\perp (T)$ below \tc, cf. Fig.~\ref{fig:ws}. We conclude that the relaxation must be dominated by the isotropic spin component, and only the coupling to the liquid has acquired a small anisotropy. 

As we move to lower $\zeta$ and approach optimal doping the systems tend to depart from the isotropic shift line with an initial slope of about 5/2, e.g. HgBa$_2$CuO$_{4+\delta}$. In particular, materials with the highest \tc appear to have the $5/2$ slope, cf.~inset in Fig.~\ref{fig:jh7}. The changes of the shifts at the lowest temperature are not well documented experimentally, and they cannot be discussed with certainty (some details are given in \cite{Haase2017}). The nuclear relaxation remains rather similar for \cperp, but the anisotropy of the relaxation changes. 

It is obvious from Fig.~\ref{fig:ws} that $K_\perp$ is nearly proportional to $1/(T_{1\perp, \parallel}T)$, however, while $1/(T_{1\perp}T)$ drops from about \SI{17}{/Ks} and \SI{25}{/Ks} to zero for both systems, respectively, the shifts have to be rescaled. For \ce{TlSr2CaCu2O$_{7-\delta}$} the shift drops from about 0.6\% to zero, cf. inset in Fig.~\ref{fig:ws}, and for \ce{YBa2Cu4O8} from about 0.1\%. From the Korringa relation one would expect $1/(T_{1\perp}T)$ of \SI{9.6}{/Ks} and \SI{0.26}{Ks}, respectively, very different values. The used scaling ratio between the two shifts in the main panel is 15/4, almost a factor of 4. 
Again, we observe a further suppression of the shifts compared to relaxation.

In a classical scenario, one expects that the relaxation governing local field fluctuations are perpendicular to the orientation of the magnetic field. Thus, in-plane fluctuations set $1/T_{1\parallel}$, while $1/T_{1\perp}$ (measured with the field in the plane) is determined by both kinds of fluctuations, i.e. parallel and perpendicular to the plane. Of course, the mean values of the shifts (proportional to $\chi_0$) do not determine their r.m.s.~averages that are set by $\chi''$ at the nuclear frequency, but they might be a good first guess for seeking a relation. For example, $K_\parallel$ is on average much larger than $K_\perp$, but there are exceptions to that rule, e.g., in terms of $(1/T_{1\perp}T)/(1/T_{1\parallel}T)$ \cite{Jurkutat2019}. In addition, we do not see a simple way to derive the special proportionality constants for $(1/T_{1\perp}T)/(1/T_{1\parallel}T)$, that hint at matrix element effects, so that we do not pursue this model any further. 

\begin{figure}[!]
\centering
\includegraphics[scale=0.22]{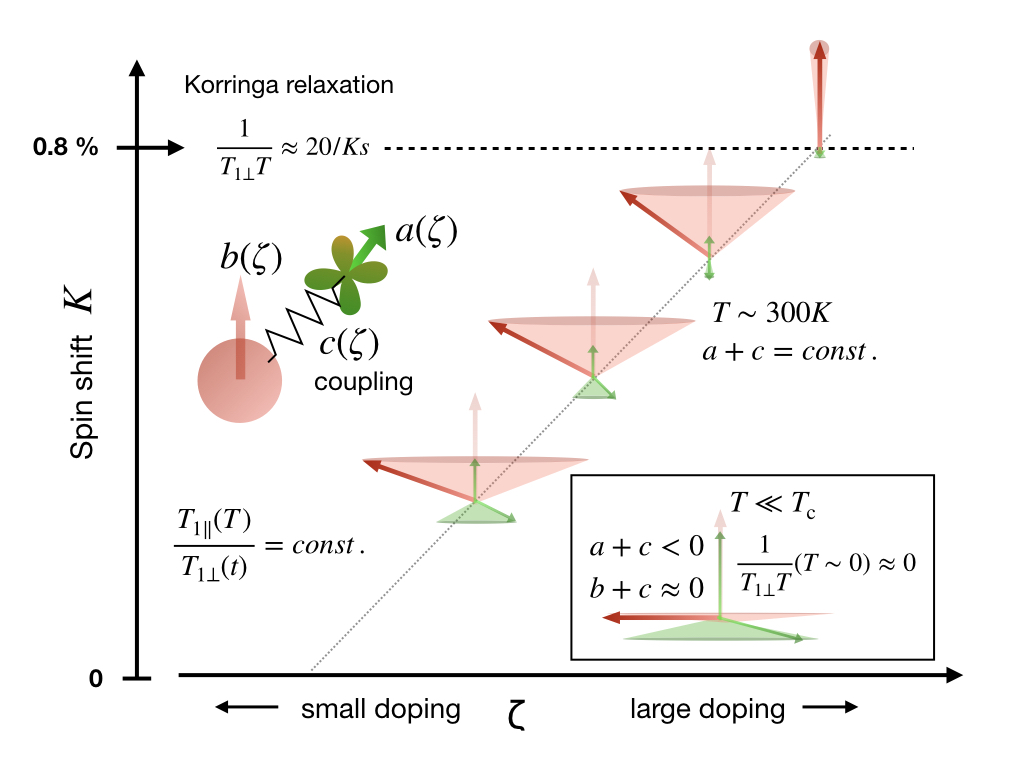}
 \caption{Shift and relaxation scenario in the cuprates: two coupled ($c(\zeta)$) electronic spins ($a(\zeta)$ and $b(\zeta)$) precess about the external magnetic field and determine the nuclear spin shift ($K(\zeta)$). The parameter $\zeta$ determines the high-\temp shifts as a function of doping and material. Near room temperature, $c(\zeta)$ is found to be temperature independent, causing isotropic changes in the NMR shifts. The largest shifts ($\sim 0.8\%$) are observed for the largest $\zeta$, in agreement with Korringa-like relaxation ($\sim \SI{20}{/Ks}$). While the relaxation is hardly affected by the coupling (only the relaxation anisotropy $(1/T_{1\perp})/(1/T_{1\parallel})$ changes), the shifts become increasingly suppressed for smaller $\zeta$, thus appear to violate the Korringa law. The coupled spins possess $s$- and $d(x^2-y^2)$-like orbital symmetry, and with the corresponding hyperfine coefficients the negative coupling explains the unexpected spin shifts observed in the cuprates. Deep in the condensed state $b+c = 0$ and relaxation disappears, but $a+c$ can be finite. The changes in $\zeta$ must be related to the pseudogap.}
\label{fig:colfig}
\end{figure}

In Fig.~\ref{fig:colfig} we illustrate a scenario of two coupled spins that we believe captures main elements observed.

There are very few systems that do \textit{not} fit the general shift scenario, among them \lsco \cite{Haase2017}. This is also true for the relaxation \cite{Jurkutat2019}, where an additional mechanism increases the relaxation above \tc, but both rates stay proportional to each other. Therefore, we also do not pursue these few outlier systems here. 

The very low-$\zeta$ materials are in general not investigated with great detail. It is known that the NMR signal can be lost, probably due to spin-glass behavior \cite{Hunt2001}. Greater material dependencies can be expected.

\subsection{Conclusions}
From literature analysis an almost universal planar Cu relaxation above and below \tc is found. It is Fermi liquid-like and changes only in terms of its anisotropy and \tc. Its doping independence rules out strong enhancement due to spin fluctuations. It is contrasted to the drastic variations of the Cu NMR shifts between different materials, as a function of doping or temperature above \tc that were reported recently \cite{Haase2017}. Shifts for materials with the highest doping obey the Korringa law when compared with their relaxation. So it must be concluded that for most cuprates that do not obey the Korringa law the shifts are suppressed, and it is not the relaxation that is enhanced.

The differences are explained with a simple two-component model that has two electronic spin components, $a(\zeta, T)$ and $b(\zeta, T)$, that depend on temperature ($T$) and a material parameter $\zeta$ that depends on doping. One of the electronic spin components, $b(\zeta,T)$, couples through an isotropic hyperfine constant, $B$, with the nuclei, while an anisotropic hyperfine constant, $A\equiv -| A_\parallel |, A_\parallel \gg A_\perp$, as known for the $3d(x^2-y^2)$ orbital is responsible for anisotropic term, $a(\zeta,T)$. A negative coupling, $c(\zeta,T)$, between both spin components, $a$ and $b$, leads to the reduction of the shifts while allowing for a largely unchanged relaxation above \tc. This negative coupling can also resolve the long standing discrepancy between calculated and presumed experimental orbital shifts.

For large $\zeta$ we find a Fermi liquid-like fluid with isotropic coupling to the nuclei, as given by the Korringa relation with shift $K(\zeta)$. As $\zeta$ decreases, $a(\zeta)$ increases, but the magnitude of the negative coupling $c(\zeta)$ suppresses the shifts (while \tc increases). Thus, $c(\zeta)$ must be related to the pseudogap. In a possible scenario $c(\zeta)$ becomes \temp dependent above \tc and causes the NMR pseudogap phenomenon, i.e., it suppresses the shifts as a function of \temp already above \tc. In this case we can conclude for the hyperfine coefficients that $A \approx - 3/5 B$.

We think it is established with NMR, now, that there is a nearly universal fluid that is Fermi liquid-like in the cuprates. This was found with NMR in 2009 \cite{Haase2009}, but also with an increasing number of other probes, e.g., \cite{Nicolas2007,Sebastian2008,Barisic2013}. Then, the most simplistic scenario suggested by our data is that the electronic spin of this liquid is coupled to the spin component in the $3d(x^2-y^2)$ orbital. Of course, the latter spin could be part of the nearly universal liquid, as well.
The term $b+c$ could be associated with quasiparticles in the nodal region of the Fermi surface \cite{Kanigel2006,Fournier2010,Proust2016,Schmidt2011}, while the term $a+c$ represents the antinodal region with perhaps antiferromagnetic properties \cite{Cilento2014,Cilento2018}. For lower values of $\zeta$, antinodal regions could be large \cite{Kaminski2015} and below \tc antiferromagnetic correlations could exist with pairing. This could explain the reduction of shift being more gradual, in comparison to overdoped samples with a smaller $k$-space region. Then, $c$ is perhaps responsible for driving the k-space anisotropy, seen by ARPES and other techniques. Neutron scattering will mostly be determined by the $a$ component and its coupling to $b$, while the response from $b$ is likely to be distributed in reciprocal space and might escape detection.

Perhaps, a Fermi liquid could reside in a separate band and inter-band coupling is responsible for the high \tc \cite{Bussmann2017}. The residual shift (that may be moments \cite{Haase2017}) could be related to time reversal symmetry breaking, but whether loop currents \cite{Varma1997} could be involved in the suppression of the shifts has to be seen. A two-component model involving hidden fermions \cite{Sakai2016,Imada2018} should relate to our findings.

Finally, we would like to mention from an NMR point of view that the evolution of the intra unit cell charge ordering that is now well documented also by NMR \cite{Reichardt2018} could be connected to the coupling scenario.

\subsection*{Acknowledgements}
We acknowledge stimulating discussions with A. Pöppl, G.V.M. Williams, A. Bussmann-Holder, M. Jurkutat, D. Dernbach, and financial support from the University of Leipzig, and the Deutsche Forschungsgemeinschaft (DFG, project 23130964). 
J.H. acknowledges the encouragement form the late C.P. Slichter to further address cuprate shifts and relaxation in numerous discussions, as well as from J. Zaanen.
\subsubsection*{Author contributions}
M.A. gathered data, made figures, helped prepare the manuscript. D.P. helped in relating the NMR findings to those obtained with other methods, discussing the contents and improving the manuscript. J.H. supplied the main concepts, wrote the manuscript and had the overall project leadership.  

\subsection*{Appendix}
A collection of abbreviations used for the various compounds is given in Table~\ref{tab}.
\begin{table}[H]
\caption{List of abbreviations with full stoichiometric formula and reference for the original data.\label{tab}}
\centering
\begin{tabular}{lll}
\textbf{Symbol}&\textbf{System}&\textbf{Ref.}\\
\hline\\[-1.50mm]
Y1248-UN-92K&		$\ce{YBa2Cu4O8}$&								\cite{Bankay1994}$^\mathrm{1}$\\
  Y1212-OP-90K		 & $\ce{YBa_2Cu_3O_{6.92}} $ &						\cite{Auler1999}$^\mathrm{1}$\\
Tl1212-OV-10K,-52K,-70K&		TlSr$_2$CaCu$_2$O$_{7-\delta}$&						\cite{Magishi1996}$^\mathrm{1}$\\
Tl2201-OV-0K,-40K,-72K&		$\ce{Tl2Ba2CuO_{6+y}}$&							\cite{Fujiwara1991,Kambe1993}$^\mathrm{1}$\\
Tl2212-OP-112K&		$\ce{Tl_2Ba_2CaCu_2O_{8-\delta}}$&							\cite{Gerashchenko1999}$^\mathrm{1}$\\
Hg1201-UN-45K,-74K&		HgBa$_2$CuO$_{4+\delta}$&						\cite{Rybicki2015}\\
Hg1201-OP-97K,-OV-85K&		HgBa$_2$CuO$_{4+\delta}$&						\cite{Rybicki2015}\\
Hg1223-OP-133K(IP)  &		HgBa$_2$Ca$_2$Cu$_3$O$_{8+\delta}$ &					\cite{Magishi1995,Julien1996}$^\mathrm{1}$\\
Hg1212-OP-127K	&$ \ce{HgBa_2CaCu_2O_{6+\delta}} $										&\cite{Itoh2017}$^\mathrm{1}$\\
Ba0223-OP-120K(OP), (IP)&		$\ce{Ba2Ca2Cu3O6(F{,}O)_2}$&				 \cite{Shimizu2011}$^\mathrm{1,2,3}$\\
Ba0212-OP102K,-OP105K&		$\ce{Ba2CaCu2O6(F{,}O)_2}$&						\cite{Shimizu2011}$^\mathrm{1}$\\
\hline\\
\end{tabular}
\vspace{-3mm}
\begin{itemize}
\item[$^1$] For the corresponding shift corrections cf. \cite{Haase2017}
\item[$^2$] OP or IP in parentheses refer to the outer and inner plane of the triple layer systems, respectively.
\item[$^3$] In Fig.~\ref{fig:jh7} the orange curve corresponds to (IP) and the yellow curve to (OP), different from Ref. \cite{Shimizu2011}. 

\end{itemize}
\end{table}

\vspace{0.5cm}

\bibliography{MANatComm.bib}   

\printindex
\end{document}